\documentclass[preprint,amsmath,amssymb,superscriptaddress,longbibliography,aps,floatfix,footinbib]{revtex4-2}
\usepackage[]{graphicx}
\usepackage{tabularx}
\usepackage[usenames,dvipsnames]{color}
\usepackage{soul}
\usepackage{bm}
\usepackage{amsmath}
\usepackage{gensymb}

\usepackage{amsfonts}
\usepackage{mathrsfs}
\usepackage{graphicx}
\usepackage{dcolumn}
\usepackage{bm}
\usepackage{color}

\usepackage[colorlinks,bookmarks=false,citecolor=blue,linkcolor=red,urlcolor=blue]{hyperref}
\usepackage{multirow}
\usepackage{physics}
\usepackage{siunitx}
\DeclareSIUnit{\barpressure}{bar}
\DeclareSIUnit\angstrom{\protect \text {Å}}

\begin{document}

\title{Frieze charge-stripes in a correlated kagome superconductor CsCr$_3$Sb$_5$}

\author{Siyu Cheng}
\affiliation{Department of Physics, Boston College, Chestnut Hill, Massachusetts 02467, USA}

\author{Keyu Zeng}
\affiliation{Department of Physics, Boston College, Chestnut Hill, Massachusetts 02467, USA}

\author{Yi Liu}
\affiliation{School of Physics, Key Laboratory of Quantum Precision Measurement of Zhejiang Province, Zhejiang University of Technology, Hangzhou 310023, P. R. China.}

\author{Christopher Candelora}
\affiliation{Department of Physics, Boston College, Chestnut Hill, Massachusetts 02467, USA}

\author{Ziqiang Wang}
\affiliation{Department of Physics, Boston College, Chestnut Hill, Massachusetts 02467, USA}

\author{Guang-Han Cao}
\affiliation{School of Physics,  Institute of Fundamental and Transdisciplinary Research, Zhejiang University, Hangzhou 310058, P. R. China}

\author{Ilija Zeljkovic}
\affiliation{Department of Physics, Boston College, Chestnut Hill, Massachusetts 02467, USA}

\maketitle

\section{Abstract}
\textbf{Kagome metals have developed into a vibrant playground for materials physics, where geometric frustration, electronic correlations and band topology come together to create a variety of exotic phenomena \cite{Wilson2024AV3Sb5Superconductors, Yin2022TopologicalSuperconductors}. Recently synthesized CsCr$_3$Sb$_5$ has provided a rare opportunity to explore unconventional superconductivity in a strongly correlated kagome system with hints of frustrated magnetism and quantum criticality \cite{Liu2024}. Using spectroscopic imaging scanning tunneling microscopy, we reveal a cascade of density wave transitions with different symmetries in bulk single crystals of CsCr$_3$Sb$_5$. In particular, we discover a new electronic state $-$ a unidirectional density wave that breaks all mirror symmetries akin to a chiral density wave, but in contrast retains a mirror-glide symmetry. We term this state a frieze charge-stripe order phase, because its symmetry properties agree with one of the fundamental frieze symmetry groups. A combination of high-resolution imaging, Fourier analysis and theoretical simulations uncovers the crucial role of sublattice degrees of freedom in forming this phase, with internal chiral textures of opposite handedness. Our experiments reveal that superconductivity in CsCr$_3$Sb$_5$ develops from a new type of a unidirectional density wave, and set the foundation for exploring electronic states with frieze symmetry groups in quantum materials. }

\section{Introduction}
Unconventional superconductivity generally occurs in the vicinity of other correlated electron phases where charge, spins or orbitals tend to form periodic spatial patterns, or density waves, that break various symmetries \cite{Fradkin2015ColloquiumSuperconductors,Stewart2011}. Canonical density waves are characterized by breaking of the translational symmetry \cite{Moncton1975StudyScattering}, and sometimes point group symmetries, such as rotational \cite{Tranquada1995} or mirror \cite{Ishioka2010chiralCDW} symmetries. Breaking of the point group symmetries can be reflected in the complex electronic textures within the unit cell. In mathematics, periodic patterns in a simple one dimensional case can be classified into seven fundamental frieze symmetry groups; each one, in addition to translation, obeys other symmetries such as certain mirror, rotation or mirror-glide reflection symmetries \cite{Morier-Genoud2015frieze}. While the periodically repeating frieze patterns naturally occur in the macroscopic world around us, and are widely used in decorative arts and architecture, realization of these symmetries by electrons at the atomic level in quantum solids has not been fully explored yet. These could however lead to the realization of new types of electronic states. For instance, a frieze pattern that breaks all mirror symmetries could give rise to a unidirectional "chiral" density wave, while the one that also preserves a mirror-glide reflection could materialize certain staggered loop-current phases.

Crystalline materials composed of atoms arranged on a corner-sharing triangular lattice, or a kagome lattice, have become an exciting platform to realize a wide array of unusual quantum states. Within this realm, the family of $A$V$_3$Sb$_5$ kagome superconductors \cite{Wilson2024AV3Sb5Superconductors, Neupert2022AVS_review} exhibits an impressive landscape $-$ various density waves \cite{Ortiz2020CVS,Jiang2021UnconventionalKV3Sb5, Zhao2021}, rotational symmetry breaking of the electronic structure \cite{Li2022RotationKV3Sb5, Wu2023, Nie2022Charge-density-wave-drivenSuperconductor,Xu2022MOKEdomains, Li2023}, topological surface states \cite{Ortiz2020CVS,Hu2022CVS_TSS} and the anomalous Hall effect \cite{Yang2020GiantKV3Sb5} $-$ with a notable absence of spin magnetism \cite{Kenney2021AbsenceSpectroscopy}. Complementary to $A$V$_3$Sb$_5$, the newly discovered Cr-based superconducting cousin CsCr$_3$Sb$_5$ is characterized by magnetic frustration and substantially stronger electron correlations \cite{Liu2024}. Combined with a highly desirable electronic flat band placed near Fermi level \cite{Liu2024, Guo2024Cr135_flatband, Li2024cr135_flatband, xieCr135, Peng2024cr135}, CsCr$_3$Sb$_5$ generated immediate excitement \cite{Sangiovanni2024}. At ambient conditions below about 55 K, CsCr$_3$Sb$_5$ undergoes a phase transition reported to be predominantly of structural origin, and superconductivity ultimately develops under pressure \cite{Liu2024}. As the material has only been recently synthesized, there is still little known about the nature of this transition, especially at the atomic length scale where insights have remained elusive. 

Here we use temperature-dependent spectroscopic-imaging scanning tunneling microscopy (SI-STM) to reveal a sequence of orthogonal unidirectional density waves in CsCr$_3$Sb$_5$: modulations with \textbf{Q$_1$}=$\frac{1}{4}$\textbf{Q$_{Bragg}$} wave vector along $\Gamma$-M emerge below 50 K and modulations with \textbf{Q$_2$}=$\frac{\sqrt 3}{8}$ $\abs{\textbf{Q$_{Bragg}$}}$ wave vector along $\Gamma$-K form below 45 K. Importantly, high-resolution imaging and Fourier analysis reveals that \textbf{Q$_2$} modulations exhibit distinct intra-unit-cell structure that breaks all in-plane mirror symmetries akin to a chiral state, but maintains a single mirror-glide symmetry distinct from known chiral density waves. As such, "handedness" that characterizes chiral density waves cannot be defined for the \textbf{Q$_2$} density wave. Since the spatial pattern can be described by one of the fundamental frieze symmetry groups, we term the newly observed unidirectional electronic state a frieze charge-stripe phase. The emergence of \textbf{Q$_2$} charge-stripes coincides with the peculiar inflection and thermal hysteresis in resistivity measurements, indicating their bulk origin. Our theoretical simulations reveal a crucial role of sublattice degrees of freedom in the formation of this state. The unusual real-space internal structure and the underlying symmetries should have profound implications on understanding of unconventional superconductivity emerging under pressure in CsCr$_3$Sb$_5$.

\section{Results}
\begin{figure}
    \centering
    \includegraphics[width = 0.65\textwidth]{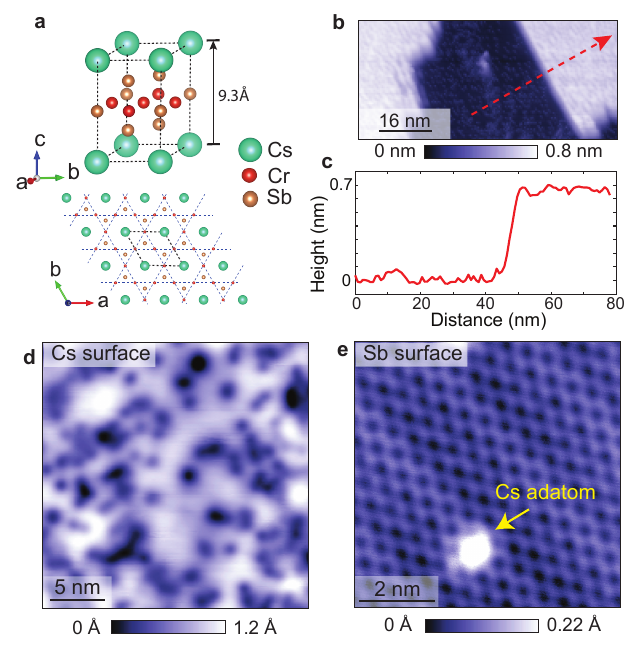}
    \renewcommand{\baselinestretch}{1}
    \caption{\textbf{Crystal structure and surface identification of CsCr$_3$Sb$_5$} - \textbf{a} 3D ball model of the CsCr$_3$Sb$_5$ crystal structure  (top panel) and the atomic structure in the $ab$-plane (bottom panel). \textbf{b} STM topograph across a step edge between two terraces: the Sb (lower) and the Cs (higher). \textbf{c} Apparent topographic height along the dashed red line in (b) showing a 6.9 $\mathring{\text{A}}$ height difference between the Cs and the Sb termination, consistent with the expected bulk structure. \textbf{d,e} Representative STM topographs of the Cs surface and the Sb surface termination, respectively. STM setup conditions: \textbf{b}, $V_{sample}$ = 1 V, $I_{set}$ = 10 pA; \textbf{d}, $V_{sample}$ = 100 mV, $I_{set}$ = 400 pA; \textbf{e}, $V_{sample}$ =300 mV, $I_{set}$ = 500 pA.}
    \label{fig:1}
\end{figure}

CsCr$_3$Sb$_5$ is a layered material with a hexagonal crystal structure, composed of Cr-Sb slabs stacked between Cs layers (Fig.~\ref{fig:1}a) \cite{Liu2024}. The kagome network of Cr atoms is located in the middle of each Cr-Sb slab, and interlaced with a hexagonal lattice of Sb atoms. We cleave bulk single crystals of CsCr$_3$Sb$_5$ in ultra-high vacuum at cryogenic temperature and immediately insert them into the STM head (Methods). Similarly to the cleavage structure of the V-based counterpart CsV$_3$Sb$_5$ \cite{Zhao2021,Liang2021}, CsCr$_3$Sb$_5$ is anticipated to cleave between the Cr-Sb slab and the Cs layer. Consistent with this expectation, STM topographs show two types of surface morphologies, the Cs layer (Fig.~\ref{fig:1}d), and the Sb layer (Fig.~\ref{fig:1}e). Large-scale STM topographs of the Cs layer are characterized by pronounced dark pits, likely a consequence of one or more missing Cs atoms. STM topographs of the Sb surface show a hexagonal lattice with a few scattered Cs atoms remaining on the Sb surface after the cleave (Fig.~\ref{fig:1}e). Using the procedure established in CsV$_3$Sb$_5$ \cite{Zhao2021}, we use the STM tip to ''sweep'' Cs atoms to the side of the field-of-view and expose a large pristine area of the Sb surface for our experiments. STM topographs reveal the step height between the Sb terrace (lower) and the Cs terrace (higher) to be 6.9 $\mathring{\text{A}}$, consistent with the distance from the lower Sb layer to the top Cs layer within one unit cell of the bulk crystal structure (Fig.~\ref{fig:1}b,c). Similar to the STM studies of all 135 materials \cite{Zhao2021, Liang2021, Li2022RotationKV3Sb5,Li2023CTB-135}, the kagome layer is never found to be directly exposed. In our experiments, we focus on imaging the Sb surface termination, which is located directly above the Cr kagome plane.

\begin{figure}
    \renewcommand{\thefigure}{1}  
    \renewcommand{\figurename}{Extended Data FIG.}
    \centering
    \renewcommand{\baselinestretch}{1}
    \includegraphics[width = 0.6\textwidth]{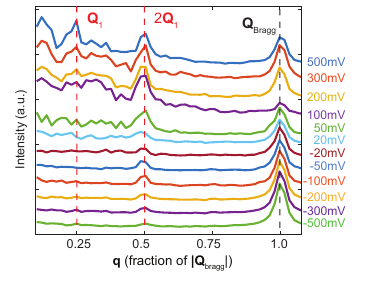}
    \caption{\textbf{Fourier transform linecuts along Q$_1$ direction.} Fourier transform linecuts, offset for clarity, along the \textbf{Q}$_1$ direction, same as those plotted in Fig.\ 2g, but in a different visual representation. STM setup conditions: $I_{set}$ = 300 pA.}
    \label{Dispersions_Q12}
\end{figure}

Low temperature STM topographs acquired at 4.5 K show pronounced signatures of unidirectionality (Fig.\ \ref{fig:2}a,c). In addition to the atomic Bragg peaks of the hexagonal lattice (denoted by diamonds Fig.\ \ref{fig:2}b), Fourier transforms (FTs) of the STM topograph show numerous other reciprocal space ($q$-space) peaks (Fig.\ \ref{fig:2}b). By a methodical examination of the FT, we can explain all these peaks as originating from two main wave vectors orthogonal to one another, \textbf{Q$_1$} = $\frac{1}{4}$\textbf{Q$_{Bragg}$} along the $\Gamma$-M direction and \textbf{Q$_2$}=$\frac{\sqrt3}{8}$ $\abs{\textbf{Q$_{Bragg}$}}$ along the $\Gamma$-K direction (red and blue circles in Fig.\ \ref{fig:2}b). All other peaks in the FT are either higher harmonics of \textbf{Q$_1$} and \textbf{Q$_2$}, \textbf{$n \times$Q$_i$} (where $n$ is an integer, and $i=$1 or 2), or the Bragg satellite peaks, \textbf{Q$^*_i$} = \textbf{Q$_{Bragg}$} $\pm$ \textbf{$n \times$Q$_i$} (dashed red and blue circles in Fig.\ \ref{fig:2}b). These can also be seen in representative FT linecuts (Fig.\ \ref{fig:2}d-f). We find that the $q$-space positions of these peaks do not change position in FTs of STM topographs or differential conductance d$I$/d$V$ maps in a wide range of biases (Fig.\ \ref{fig:2}g,h,i, Extended Data Fig.~\ref{Dispersions_Q12}). On this basis, we can attribute these to charge ordering states with wave vectors \textbf{Q$_1$} and \textbf{Q$_2$}.

\begin{figure}
    \centering
    \includegraphics[width = \textwidth]{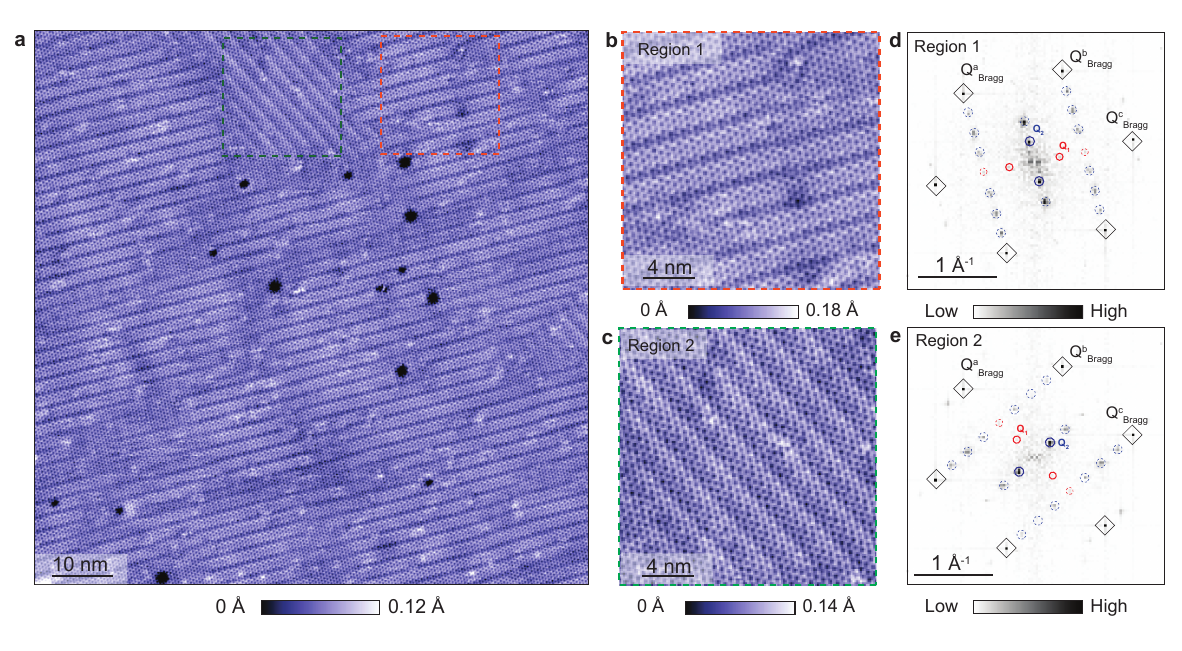}
    \renewcommand{\baselinestretch}{1}
    \caption{\textbf{Imaging of domains with different charge-stripe orientation.} \textbf{(a)} STM topograph encompassing multiple domains, two of which are enclosed by dashed green and orange squares. This rules out that unidirectionality observed is a consequence of an aniostropic STM tip. \textbf{(b,c)} Zoom-ins on the two domains denoted by the squares in (a), and \textbf{(d,e)} their associated Fourier transforms (FTs). Atomic Bragg peaks \textbf{Q}$^i_{Bragg}$ ($i$=$a,b,c$) are enclosed in black diamonds. Two main wave vectors \textbf{Q}$_1$ and \textbf{Q}$_2$ and enclosed in red and blue solid circles, respectively. Higher order harmonics and satellite peaks are enclosed in dashed circles. It can be seen that both \textbf{Q}$_1$ and \textbf{Q}$_2$ rotate together, and are orthogonal to one another in different domains. STM setup conditions: \textbf{a,b,c}, $V_{sample}$ = -500 mV, $I_{set}$ = 300 pA.  }
    \label{fig:3}
\end{figure}

We proceed to examine large-scale STM topographs, which show nanoscale domains (Fig.\ \ref{fig:3}a). Within each domain, prominent stripe modulations are rotated by 60 degrees with respect to one another across a domain wall (Fig.\ \ref{fig:3}b,c). Size of domains in our samples varies from few to tens of nanometers. The observation of domains oriented along different directions within the same field-of-view and acquired with the same STM tip conclusively demonstrates that observed electronic unidirectionality is not a consequence of an anisotropic tip. Moreover, by inspecting STM topographs and FTs within individual domains, we can clearly see that both \textbf{Q$_1$} and \textbf{Q$_2$} are rotated by 60 degrees in domains oriented in different directions (Fig.\ \ref{fig:3}d,e), thus demonstrating an intimate symmetry connection between the two wave vectors.

\begin{figure}
    \renewcommand{\thefigure}{2}  
    \renewcommand{\figurename}{Extended Data FIG.}
    \centering
    \renewcommand{\baselinestretch}{1}
    \includegraphics[width = 0.8\textwidth]{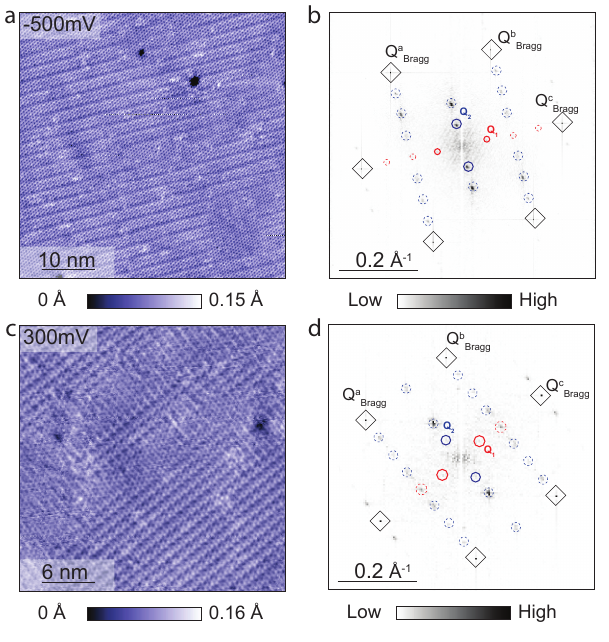}
    \caption{\textbf{Reproducibility of observations of different density waves.} \textbf{(a,b)} STM topograph and corresponding Fourier transform obtained on sample 1, also from main text. \textbf{(c,d)} STM topograph and corresponding Fourier transform obtained on sample 2. Data on the two samples was obtained using two different STM tip wires. STM setup conditions: \textbf{a}, $V_{sample}$ = -500 mV, $I_{set}$ = 200 pA; \textbf{b}, $V_{sample}$ = 300 mV, $I_{set}$ = 200 pA.}
    \label{repeatbility_of_Q12}
\end{figure}

To gain further insight into the relationship between \textbf{Q$_1$} and \textbf{Q$_2$}, we perform STM experiments as a function of temperature (Fig.~\ref{fig:4}). At 50 K, STM topographs only show the three pairs of atomic Bragg peaks (Fig.~\ref{fig:4}d). Below 49 K however, stripe modulations associated with \textbf{Q$_1$} begin to emerge, while the modulations with \textbf{Q$_2$} wave vector remain absent (Fig.~\ref{fig:4}a,e). This suggests the onset temperature of \textbf{Q$_1$} modulations to be $T_1 \approx 50 K$. Lowering the temperature further, we find that modulations with \textbf{Q$_2$} wave vector also begin to appear below $T_2 = 45$ K and coexist with \textbf{Q$_1$} modulations (Fig.~\ref{fig:4}b,f,g). The appearance of the two types of density waves by lowering the temperature can also be visualized by the waterfall plot of the FT linecuts (Fig.~\ref{fig:4}h,i). Both modulations remain present towards the base temperature of 4.5 K. We have verified the existence of both types of modulations on multiple samples with different STM tips (Extended Data Fig.~\ref{repeatbility_of_Q12}). This measurement uncovers close, yet clearly distinct, temperature scales of the two charge ordering states in this material.

\begin{figure}
    \renewcommand{\thefigure}{4}  
    \renewcommand{\figurename}{Extended Data FIG.}
    \centering
    \renewcommand{\baselinestretch}{1}
    \includegraphics[width = \textwidth]{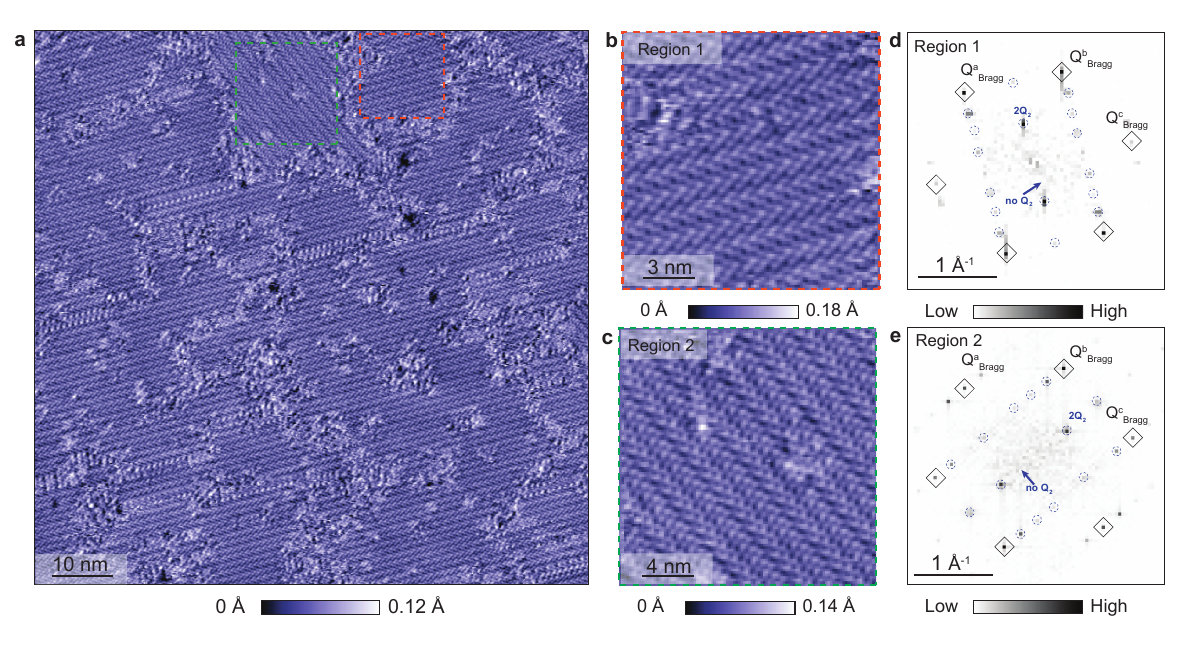}
    \caption{\textbf{Additional data on an area encompassing multiple domains.} \textbf{(a)} Large STM topograph showing multiple domains. \textbf{(b,c)} Smaller zoom-ins on the two regions enclosed by dashed red and green rectangles, and \textbf{(d,e)} associated Fourier transforms. Both regions show the absence of \textbf{Q}$_2$ but strong Bragg satellite peaks. STM setup condition: (a) $V_{sample}$ = 50 mV, $I_{set}$ = 300 pA.}
    \label{domain_Q2_rotation}
\end{figure}

\begin{figure}
    \renewcommand{\thefigure}{6}  
    \renewcommand{\figurename}{Extended Data FIG.}
    \centering
    \renewcommand{\baselinestretch}{1}
    \includegraphics[width = \textwidth]{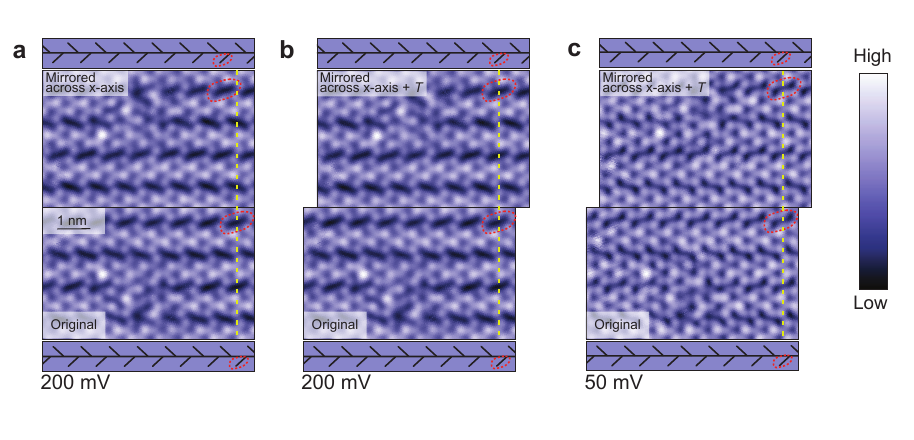}
    \caption{\textbf{Visualizing mirror-glide symmetry.} \textbf{(a)} STM topograph (lower half) and its mirror image along the $x$-axis (upper half). It can be seen that the arrow-like pattern has an abrupt transition between the two images when viewed from up to down (see for example the white dashed circles and the yellow dashed line). \textbf{(b)} The same two topographs from (a) but with the mirror image (upper one) offset horizontally by $\cos{\frac{\pi}{6}}$ of the lattice constant $a_0$. It can be seen that now the arrow-like pattern is aligned up between the original image and the transformed image. \textbf{c} Same as (b) but for a different STM bias. STM setup conditions: \textbf{a-c}, $I_{set}$ = 100 pA.}
    \label{mirror_glide}
\end{figure}

\textbf{Q$_2$} charge-stripe phase exhibits several remarkable features. First, from the symmetry perspective, the state breaks all in-plane mirror symmetries akin to a chiral charge density wave (Fig.~\ref{fig:5}a). However in contrast, it retains a mirror-glide symmetry (Fig.~\ref{fig:5}b) $-$ a reflection with respect to the $x$-axis and a translation along the $x$-axis maps the pattern back into itself (Fig.~\ref{fig:5}b, Extended Data Fig.~\ref{mirror_glide}). As such, global handedness of the structure cannot be defined. We term this new type of a unidirectional density wave a frieze charge-stripe phase, based on its similarity to one of the fundamental one-dimensional frieze symmetry groups characterized by a mirror-glide reflection symmetry (inset in Fig.~\ref{fig:5}a). Second, high-resolution STM topograph reveals that dark zig-zag electronic features connect two in-plane Sb atoms through a Cr atom. As such, at least 4 sublattices are needed to describe the state: 3 Cr atoms that compose the kagome lattice and one in-plane Sb atom. Third, the state exhibits an unconventional form factor reflected in the unusual variations in the amplitudes of different FT peaks (Fig.~\ref{fig:5}c). In particular, the main \textbf{Q$_2$} harmonic is nearly completely absent around Fermi energy and at positive bias (Fig.~\ref{fig:2}h,i), while its Bragg satellite peak \textbf{Q$^*_2$} remains strong (Fig.~\ref{fig:5}c). The suppression of \textbf{Q$_2$} is not an artifact of tip symmetry as this can be observed for differently oriented domains using the same STM tip (Extended Data Fig.~\ref{domain_Q2_rotation}). Lastly, despite the suppression of the main harmonic \textbf{Q$_2$}, higher harmonics 2\textbf{Q$_2$} and 4\textbf{Q$_2$} are particularly prominent (Fig.~\ref{fig:5}c), which has not been observed in other density wave systems.  

To understand the formation of the \textbf{Q$_2$} charge-stripes, we develop a theoretical model taking into account 4 atoms in the kagome layer in this system: 3 Cr atoms that comprise the kagome net and the Sb atom in the center of each hexagon. Different densities on Cr and Sb atoms can be described by an intracell density wave with wave vectors equal to the reciprocal lattice vectors $\textbf{Q}^\alpha_{ Bragg}$: $\Delta_0(\textbf{r}_i) = \rho \sum_\alpha \mathrm{cos}(\textbf{Q}^\alpha_{ Bragg} \cdot \textbf{r}_i)$, $\alpha=a,b,c$, $i=\text{Sb}, \text{Cr}_1,\text{Cr}_2,\text{Cr}_3$. The emergence of the \textbf{Q$_2$} density wave can be described by $\Delta_{\text{CDW}}(\textbf{r}_i) = \sum_{n,\alpha} \rho_{n,\alpha} \mathrm{cos}(n\left[\textbf{Q}_2+\textbf{Q}^\alpha_{ Bragg}\right] \cdot \textbf{r}_i + \theta)$, where $n=1,2,3,4$ accounts for the basic and the higher harmonics at wave vectors $n\textbf{Q}_2$, and $\theta$ is the overall phase shift of the density waves with respect to $\Delta_0$ (see Methods for more details). The simulation (Fig.~\ref{fig:5}d) presents an excellent visual agreement with our data (Fig.~\ref{fig:5}a). It obeys all the preserved and broken symmetries observed in the experimental data, and reveals chiral textures of opposite handedness produced by the different $\rho_{n,\alpha}$ along different crystalline directions, and the mirror-glide symmetry of the frieze group for $\theta=\pi$ (Fig.~\ref{fig:5}e). It also captures the suppression of the main \textbf{Q$_2$} harmonic and the enhancement of the satellite peak \textbf{Q$^*_2$} (Fig.~\ref{fig:5}f). This occurs due to a destructive sublattice interference, more easily seen in a simple scenario shown in Extended Data Fig.~\ref{pedagogical_models}b,d, similarly to the suppression of main harmonics related to the density wave in the pseudogap phase of cuprates \cite{Fujita2014d-form}.

\section{Discussion}
\begin{figure}
    \centering
    \includegraphics[width = \textwidth]{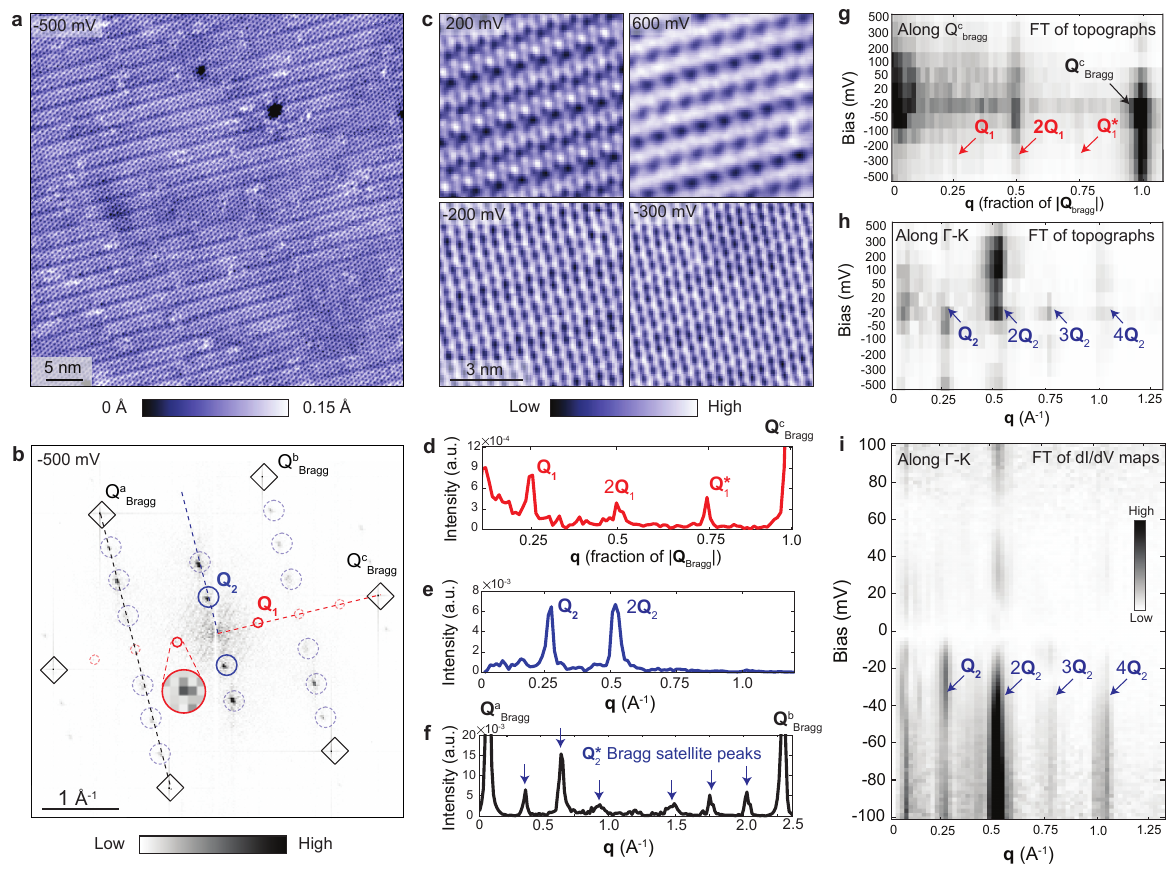}
    \renewcommand{\baselinestretch}{1}
    \caption{\textbf{Identifying different types of charge modulations at low temperature.} \textbf{(a)} STM topograph of the Sb surface showing stripe-like modulations in a single domain, and \textbf{(b)} associated Fourier transform (FT). Atomic Bragg peaks \textbf{Q}$^i_{Bragg}$ ($i$=$a,b,c$) are enclosed in black diamonds. Two main wave vectors \textbf{Q}$_1$ and \textbf{Q}$_2$ and enclosed in red and blue circles, respectively. Higher order harmonics and satellite peaks are enclosed in dashed circles. Both \textbf{Q}$_1$ and \textbf{Q}$_2$ are unidirectional and perpendicular to one another. \textbf{(c)} Smaller STM topographs showing the modulations at different STM biases. FT linecuts from panel (b) along: \textbf{(d)} dashed line showing \textbf{Q}$_1$ and associated peaks, \textbf{(e)} dashed blue line showing \textbf{Q}$_2$ and higher harmonics, and \textbf{(f)} dashed black line showing all the \textbf{Q}$_2$ Bragg satellite peaks. \textbf{(g)} Waterfall plot of FT linecuts of STM topographs along the red dashed line in (b) showing the absence of dispersion of \textbf{Q}$_1$-related peaks. \textbf{(h)} Waterfall plot of FT linecuts of STM topographs along the blue dashed line in (b) showing the absence of the dispersion of \textbf{Q}$_2$-related peaks. \textbf{(i)} Waterfall plot of FT linecuts of d$I$/d$V$ maps along the blue dashed line in (b) showing the absence of dispersion of \textbf{Q}$_2$-related peaks. STM setup conditions: \textbf{a}, $V_{sample}$ = -500 mV, $I_{set}$ = 200 pA; \textbf{c}, $I_{set}$ = 50 pA; \textbf{g,h}, $I_{set}$ = 300pA; \textbf{i}, $V_{sample}$ = 100mV, $I_{set}$ = 300pA, $V_{exc}$ = 2mV.}
    \label{fig:2}
\end{figure}

Our work provides the first atomic-scale visualization of the electronic structure of the newly discovered correlated kagome superconductor CsCr$_3$Sb$_5$. Below $T_1 \approx 50$ K, the system goes through a phase transition that breaks both the translational and rotational symmetries of the crystal structure, by the formation of a commensurate unidirectional density wave with \textbf{Q$_1$}=$\frac{1}{4}$\textbf{Q$_{Bragg}$} wave vector along $\Gamma$-M. \textbf{Q$_1$} modulations imaged here by STM are identical to those seen by X-ray diffraction measurements of the same CsCr$_3$Sb$_5$ samples at approximately the same temperature \cite{Liu2024}, thus demonstrating their bulk nature. The transition has a strong structural component \cite{Liu2024}, which is also consistent with our STM experiments $-$ the electronic signal associated with \textbf{Q$_1$} is relatively weak compared to other FT peaks (Fig.~\ref{fig:2}d,e), and it is difficult to resolve close to the Fermi level (Fig.~\ref{fig:2}g,h). 

\begin{figure}
    \renewcommand{\thefigure}{4}  
    \renewcommand{\figurename}{FIG.}  
    \centering
    \includegraphics[width = \textwidth]{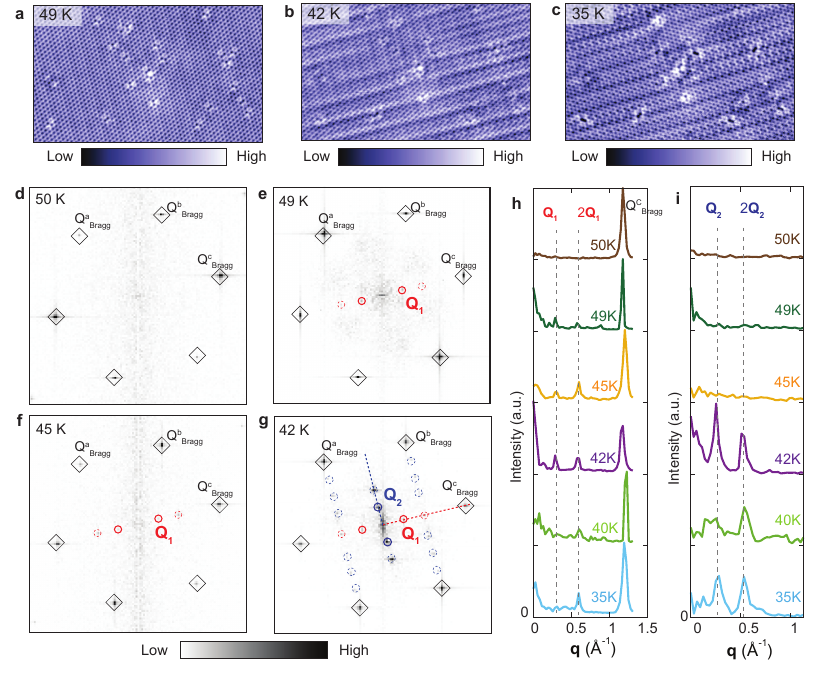}
    \renewcommand{\baselinestretch}{1}
    \caption{\textbf{Disentangling density wave transition temperatures through spectroscopic imaging STM.} (a-c) STM topographs over an identical field of view at (a) 49 K, (b) 42 K, and (c) 35 K. Only \textbf{Q$_1$} stripes are visible in (a), while both \textbf{Q$_1$} and \textbf{Q$_2$} charge stripes can be seen in (b,c). (d-g) Fourier transforms of STM topographs encompassing the sample region in (a-c), showing the emergence of Fourier transform (FT) peaks associated with \textbf{Q$_1$} and \textbf{Q$_2$} at different temperatures. FT linecuts as a function of temperature along: (h) red line in (g) through the \textbf{Q$_1$} peak, and (i) blue dashed line in (g) through \textbf{Q$_2$} peak. STM setup conditions:  \textbf{a-g}, $V_{sample}$ = -500 mV, $I_{set}$ = 50 pA.}
\label{fig:4}
\end{figure}

\begin{figure}
    \renewcommand{\thefigure}{5}  
    \renewcommand{\figurename}{Extended Data FIG.}
    \centering
    \renewcommand{\baselinestretch}{1}
    \includegraphics[width = 0.6\textwidth]{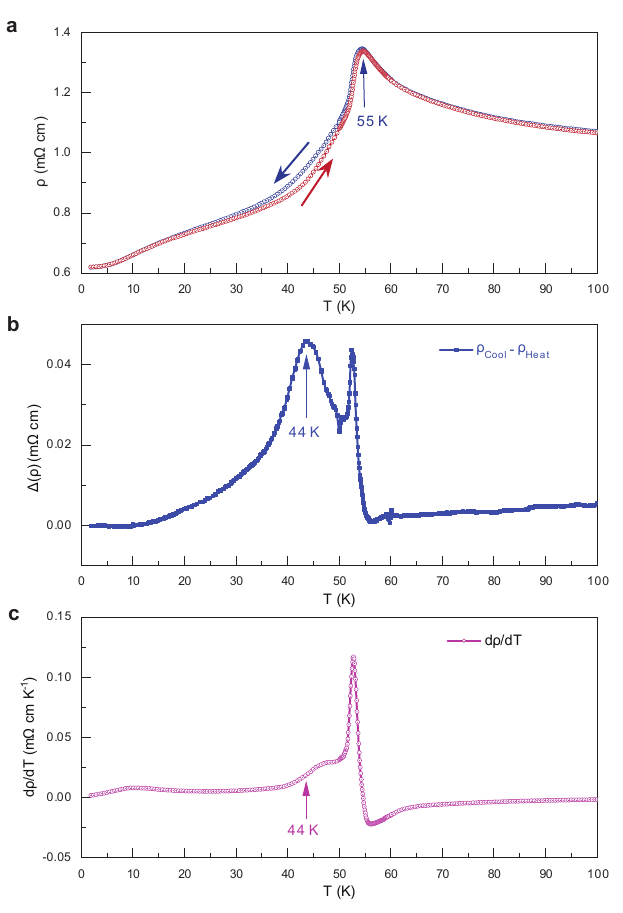}
    \caption{\textbf{Temperature-dependent resistivity measurements of CsCr$_3$Sb$_5$.} \textbf{(a)} Resistivity as function of temperature upon cooling (blue) and warming up (red) showing a thermal hysteresis and an inflection point around 45 K. \textbf{(b)} Difference between the two curves in (a) showing a peak at 44 K. \textbf{(c)} Derivative of resistivity with respect to temperature, showing an inflection point seen by eye in (a).}
    \label{resistivity_45K_kink}
\end{figure}

Below $T_2 \approx 45$ K, we discover that the system breaks an additional translational symmetry, by the emergence of a unidirectional density wave with wave vector \textbf{Q$_2$} orthogonal to \textbf{Q$_1$}. It has a strong electronic component, as evidenced by the Fourier components that are generally orders of magnitude higher in amplitude compared to that of \textbf{Q$_1$} (Fig.~\ref{fig:2}b,d,e, Fig.~\ref{fig:4}h,i), and strikingly different as a function of bias (Fig.~\ref{fig:2}c). Interestingly, although no anomalies in bulk magnetization and specific heat measurements are observed around 45 K \cite{Liu2024}, we find that electrical resistivity as a function of temperature shows an inflection point and an unusual thermal hysteresis at this temperature (Extended Data Fig.~\ref{resistivity_45K_kink}). This provides strong evidence for the bulk nature of this lower-temperature transition. We note that \textbf{Q$_2$} is yet to be detected in X-ray diffraction measurements, possibly due to the fact that the experiments thus far were performed down to 40 K temperature only \cite{Liu2024}, which we show is at the cusp of the formation of this state. It is also possible that \textbf{Q$_2$} modulations are accompanied by a weaker structural modulation given their strong electronic fingerprint detected in STM and electrical transport, and thus would be more challenging to detect in diffraction measurements. 

By uncovering microscopic signatures of density waves in CsCr$_3$Sb$_5$, our work also sets the stage for investigating whether there exist spin or unconventional orbital components associated with the orders. Theoretical possibilities put forth include for instance antiferromagnetic ordering \cite{Liu2024} and a 4 $\times$ 2 spin-density wave with an altermagnetic state \cite{Xu2023}. These can be possibly pursued in, for example, spin-polarized STM experiments \cite{Enayat2014SPSTM, Zhao2019SPSTM}. Given the gradual pressure suppression of \textbf{Q$_1$} and its ultimate coexistence with superconductivity \cite{Liu2024}, it is possible that \textbf{Q$_2$} electronic stripes also coexist with superconductivity at higher pressure \cite{Liu2024}, which can in principle give rise to a secondary pair density modulation phase with \textbf{Q$_2$} vector \cite{Ruan2018cupratesPDW,Liu2021NbSe2PDW}. In this scenario, the pair density modulation could inherit the unusual form factor, leading to the unidirectional superconductivity modulations with chiral sub-textures. This possibility will be of high interest to explore in subsequent work. Overall, our comprehensive atomic-scale, energy-resolved imaging of CsCr$_3$Sb$_5$ sets the foundation for understanding frieze symmetry charge-stripe superconductors.

\begin{figure}
    \renewcommand{\thefigure}{5}  
    \renewcommand{\figurename}{FIG.}  
    \centering
    \includegraphics[width = \textwidth]{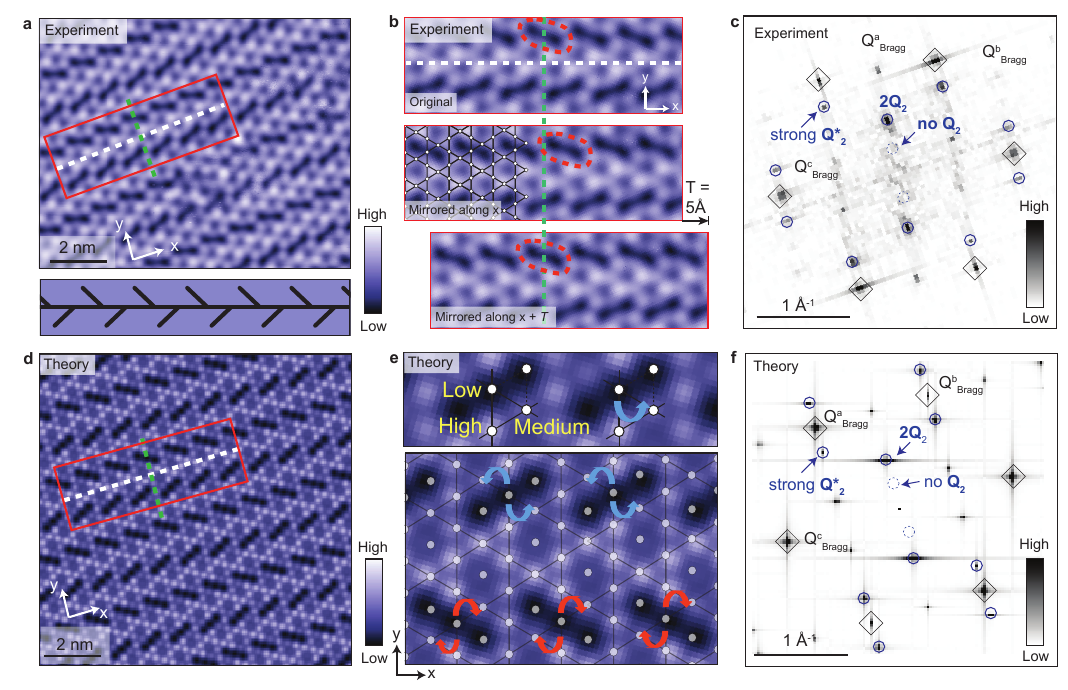}
    \renewcommand{\baselinestretch}{1}
    \caption{\textbf{Intra-unit-cell structure of the Q$_2$ density wave.} (a) Experimental STM topograph showing that all mirror symmetries are broken (dashed lines). The bottom inset shows one of the fundamental frieze symmetry groups, which matches the data outlined by the red rectangle in (a). (b) Zoom-in on the red rectangle in (a) (top), its mirror image with respect to the $x$-axis (middle), which is then shifted by about 5 $\mathring{\text{A}}$ along the $x$-axis to obtain the original pattern back (bottom). Lattice registry is obtained from Extended Data Fig.~\ref{lattice_on_topo}. (c) FT of (a). (h) Simulated topograph using the four sublattice model (Methods), which presents an excellent match to data in (a). (e) Zoom-in on small regions of the simulated topograph showing chiral sub-textures of adjacent kagome triangles within each dark stripe, denoted by different arrows. (f) FT of (d). STM setup conditions:  \textbf{a,b}, $V_{sample}$ = 100 mV, $I_{set}$ = 100 pA.}
\label{fig:5}
\end{figure}

\begin{figure}
    \renewcommand{\thefigure}{8}  
    \renewcommand{\figurename}{Extended Data FIG.}
    \centering
    \renewcommand{\baselinestretch}{1}
    \includegraphics[width = 0.9\textwidth]{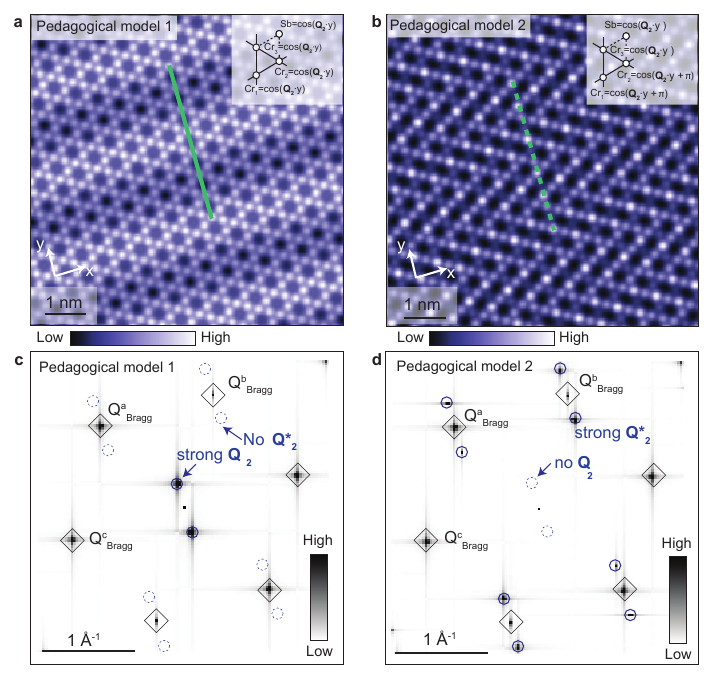}
    \caption{\textbf{Theoretical model with simplified conditions for pedagogical purposes.} We work with the same model for the \textbf{Q$_2$} density wave from the main text: $\Delta_{\text{CDW}}(\textbf{r}_i) = \sum_{n,\alpha} \rho_{n,\alpha} \mathrm{cos}(n\left[\textbf{Q}_2+\textbf{Q}^\alpha_{Bragg}\right] \cdot \textbf{r}_i + \theta)$. (a) Simulated topograph with a simple modulation $\Delta_{\text{CDW}}(\textbf{r}_i) =\mathrm{cos}(\textbf{Q}_2 \cdot \textbf{r}_i)$, where $\textbf{Q}^\alpha_{Bragg}$ is not introduced as there is no sublattice phase difference. The $\textbf{Q}_2$ peak appears around the FT center, and the modulation is unidirectional. The mirror symmetries along the $y$-axis and along the $x$-axis remain intact. (b) Simulated topograph for $n$=1 and $\theta$=0, as the expression becomes $\Delta_{\text{CDW}}(\textbf{r}_i) = \sum_{\alpha=a,b}\rho_{\alpha} \mathrm{cos}(\left[\textbf{Q}_2+\textbf{Q}^\alpha_{Bragg}\right] \cdot \textbf{r}_i)=\sum_{\alpha=a,b}\rho_{\alpha} \mathrm{cos}(\textbf{Q}_2\cdot \textbf{r}_i+\phi_\alpha)$. The term $\phi_\alpha=\textbf{Q}^\alpha_{Bragg} \cdot \textbf{r}_i$ brings different phases to the four sublattices, in the order of $\text{Sb}, \text{Cr}_1,\text{Cr}_2,\text{Cr}_3$. For $\alpha=a$, $\phi_a = 0, \pi, \pi, 0$,  for $\alpha=b$, $\phi_b = 0, 0, \pi, \pi$. The inset in (b) only shows the modulation for $\alpha=a$. Importantly, the different phases $\phi_\alpha$ within each sublattices result in a complete annihilation of the $\textbf{Q}_2$ peaks around the FT center. For $\alpha=c$, the mirror symmetries along the $y$-axis remain because $\textbf{Q}^c_{Bragg}$ is perpendicular to $\textbf{Q}_2$. If $\alpha=a$ or $\alpha=b$, the density wave will break the mirror symmetries along the $y$-axis. If these mirror planes mismatch with the mirror planes of the kagome lattice by a phase shift $\theta$, then all mirror symmetries are broken. There are also inversion symmetries at the nodes of the wave, which can be preserved by overlapping with kagome inversion centers at the sites, for example when $\theta=\pi/8$.\\}
    \label{pedagogical_models}
\end{figure}

Although kagome metals already exhibit an abundance of different charge density wave vectors and morphologies, a density wave with a \textbf{Q$_2$} wave vector has not been reported or explored theoretically. Unlike a prototypical unidirectional density wave that still preserves mirror symmetries along, and perpendicular to, its wave vector (Extended Data Fig.~\ref{pedagogical_models}a,c), unidirectional modulations associated with \textbf{Q$_2$} break all in-plane mirror symmetries but maintain a mirror-glide reflection. Our theoretical modeling suggests that this could be explained by intracell density waves in a system with three kagome sublattices and the in-plane Sb atom creating chiral textures with staggered handedness (Fig.~\ref{fig:5}d,e). In sharp contrast to chiral charge density waves that have been studied in select other systems that are tri-directional \cite{Ishioka2010chiralCDW, Jiang2021UnconventionalKV3Sb5}, \textbf{Q$_2$} charge stripes here are unidirectional and maintain a mirror-glide reflection. So from the symmetry perspective, \textbf{Q$_2$} frieze charge-stripes discovered here are fundamentally different. They can be described by a frieze symmetry group that obeys a mirror-glide reflection (Fig.~\ref{fig:5}a, inset). By adjusting the phases in our theoretical model describing this state, one can in principle create charge-stripes that further break the mirror-glide and/or inversion symmetries. This could generate for example the first unidirectional chiral phase described by a different frieze symmetry group, which could be realized in related materials. Overall, our experiments provide a foundation for exploring electronic states with frieze symmetry formalism in quantum materials.\\
 

\noindent{\bf Author contributions}\\

S.C. performed STM measurements with the help from C.C. Y.L. and G-H.C. synthesized the bulk single crystals. K.Z., S.C. performed theoretical calculations under the supervision of Z.W. and I.Z. I.Z., S.C. and Z.W wrote the paper with the input from all the authors. I.Z. supervised the project.\\
\\
\noindent{\bf Methods}\\
\indent \textit{\textbf{Sample growth:}} Single crystals of CsCr$_3$Sb$_5$ were grown by a self-flux growth method using CsSb-CsSb$_2$ mixture as the flux \cite{Liu2024}. High purity Cs (99.9\%), Cr (99.95\%) and Sb(99.99\%) were weighed in a molar ratio of Cs : Cr : Sb = 9 : 2 : 18, with a total mass of approximately 5 g. The mixture was loaded into an alumina crucible, sealed in a Ta tube and then jacketed in a silica ampule. The sample-loaded assembly was heated to 1173 K over 18 hours. After maintaining this temperature for 24 hours, the sample was allowed to cool to 873 K at a rate of 2 K/hour. The flux was subsequently removed by washing with ethanol. Plate-like crystals with size up to 0.70 $\times$ 0.70 $\times$ 0.05 mm$^3$ were harvested.

\textit{\textbf{STM experiments: }}Samples and the cleave rods were attached to the sample holder using conducting epoxy EPO-TEK H20E and cured at 175 \textdegree C for about 20 min. We cold-cleaved the crystals in UHV at a cryogenic temperature (approximately few tens of Kelvin) and quickly insert them into the STM head. STM data was acquired using a customized Unisoku USM1300 microscope. STM tips used were home-made, chemically-etched tungsten tips.

\textit{\textbf{Theoretical simulation: }} The model in Fig.~\ref{fig:5} takes into account 4 sublattices in the original kagome-Sb layer: three Cr atoms and one Sb atom. Different densities on the Cr and Sb atoms can be described by an intracell density wave with wavevectors equal to the reciprocal lattice vectors $\textbf{Q}^\alpha_{Bragg}$: $\Delta_0(\textbf{r}_i) = \rho
_0\sum_\alpha \mathrm{cos}(\textbf{Q}^\alpha_{Bragg} \cdot \textbf{r}_i)$, $\alpha=a,b,c$, $i=\text{Sb}, \text{Cr}_1,\text{Cr}_2,\text{Cr}_3$. This density wave keeps all kagome lattice symmetries when the density modulations $\rho$ along $\textbf{Q}^\alpha_{Bragg}$ are equal. To describe the density wave with the experimentally observed FT peaks, we use the model: $\Delta_{\text{CDW}}(\textbf{r}_i) = \sum_{n,\alpha} \rho_{n,\alpha} \mathrm{cos}(n\left[\textbf{Q}_2+\textbf{Q}^\alpha_{Bragg}\right] \cdot \textbf{r}_i + \theta)$, where $n=1,2,3,4$ accounts for the basic and the higher harmonics of the $n\textbf{Q}_2$ density waves and $\theta$ is an overall phase shift of the density wave with respect to $\Delta_0$. The contribution from n=odd terms breaks all mirror reflection symmetries locally when $\rho_{n, a}=-\rho_{n, b}$ and $\theta=\pi/8$, but forms the mirror-glide symmetry. Moreover, the $n$=odd $n\textbf{Q}_2$ peaks are absent in the FT spectrum due to sublattice interference. On the other hand, for 
$n$=even $n\textbf{Q}_2$ peaks, the term $n\phi_\alpha=n\textbf{Q}^\alpha_{Bragg} \cdot \textbf{r}_i$ brings $0, 0, 2\pi, 2\pi$ phases to the four sublattices. Therefore, the density waves are effectively $\Delta_{\text{CDW}}(\textbf{r}_i) =\rho_n\mathrm{cos}(n\textbf{Q}_2 \cdot \textbf{r}_i)$, and the FT peaks appear around the $\Gamma$ point without phase canceling; these also preserve inversion and mirror-glide symmetries. Taking into account $n=1,2,3,4$ density waves, we obtain the full simulation shown in Fig.~\ref{fig:5}d with all symmetries observed in the experiments.\\

\begin{table}
    \centering
    \begin{tabular}{|c|c|c|c|} \hline  
        & a & b & c\\ \hline 
        $\rho_{1,\alpha}$&  0.5&  -0.5&  0.0\\ \hline  
        $\rho_{2,\alpha}$&  0.0&  -1.0&  0.0\\ \hline 
        $\rho_{3,\alpha}$&  0.2&  -0.2&  0.0\\ \hline
        $\rho_{4,\alpha}$&  0.0&  -0.3&  0.0\\ \hline
    \end{tabular}                         
    \caption{Values of density wave intensities. $\rho_0=- 0.75$}
    \label{simparam}
\end{table}




\noindent{\bf Data availability}\\
The data that support the findings of this study are available from the corresponding authors upon reasonable request.\\
\\
\noindent{\bf Code availability}\\
The code that supports the findings of the study is available from the corresponding authors upon reasonable request.\\
\\
\noindent{\bf Correspondence and requests for materials} should be addressed to \textit{ghcao@zju.edu.cn}, \textit{wangzi@bc.edu} and \textit{ilija.zeljkovic@bc.edu}.\\
\\
\noindent{\bf Competing financial interests}\\
The authors declare no competing financial interests.\\

\bibliographystyle{unsrt}
\bibliography{kagome.bib}


\newpage

\begin{figure}
    \renewcommand{\thefigure}{3}  
    \renewcommand{\figurename}{Extended Data FIG.}
    \centering
    \renewcommand{\baselinestretch}{1}
    \includegraphics[width = 0.9\textwidth]{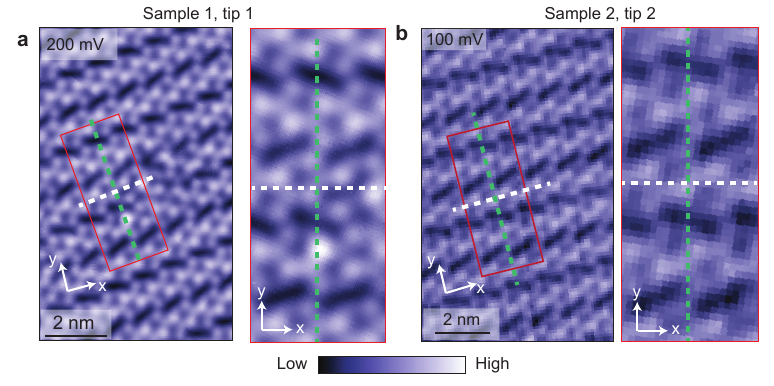}
    \caption{\textbf{Reproducibility of mirror symmetry breaking.} \textbf{(a,b)} STM topographs (left) and zoomed-in regions (right). Dashed green and white lines denote mirror symmetries that are broken. Topograph in (a) was taken on sample 1 using tip wire 1, and (b) is acquired on a different sample with a different STM tip. STM setup conditions: \textbf{a}, $V_{sample}$ = 200 mV, $I_{set}$ = 100 pA; \textbf{b}, $V_{sample}$ = 100 mV, $I_{set}$ = 200 pA.}
    \label{repeatbility_of_RSB_breaking}
\end{figure}

\begin{figure}
    \renewcommand{\thefigure}{7}  
    \renewcommand{\figurename}{Extended Data FIG.}
    \centering
    \renewcommand{\baselinestretch}{1}
    \includegraphics[width = \textwidth]{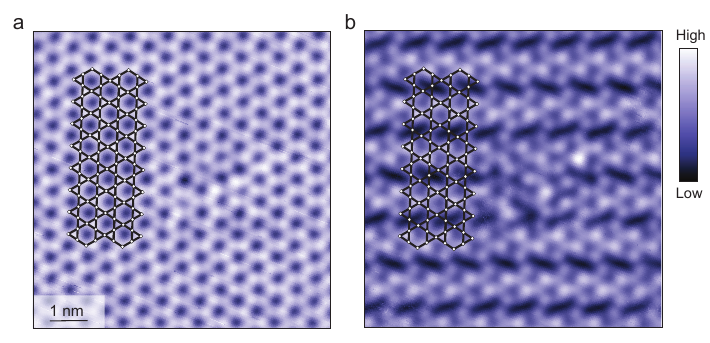}
    \caption{\textbf{Lattice Registry.} \textbf{(a,b)} STM topographs over an identical region of the sample, acquired in back-to-back scans. We can identify the lattice structure in (a) where morphology is similar to other kagome materials in the 135 family, and superimpose it in the topograph in (b). STM setup condition: \textbf{a}, $V_{sample}$ = -200 mV, $I_{set}$ = 100 pA; \textbf{b}, $V_{sample}$ = 200 mV, $I_{set}$ = 100 pA.}
    \label{lattice_on_topo}
\end{figure}

\end{document}